\documentclass[]{spie}  %>>> use for US letter paper
\usepackage{amsmath,amsfonts,amssymb}
\usepackage{graphicx}
\usepackage{float}
\usepackage[colorlinks=true, allcolors=blue]{hyperref}
\usepackage{siunitx}
\usepackage{caption}
\usepackage{subcaption}
\usepackage[final]{pdfpages}

\title{POLAR-2: A large scale gamma-ray polarimeter for GRBs}

\author[a]{J. Hulsman}
\author[a]{N. de Angelis}
\author[d]{J. M. Burgess}
\author[a]{F. Cadoux}
\author[d]{J. Greiner}
\author[a]{M. Kole}
\author[b]{H. Li}
\author[c]{S. Mianowski}
\author[c]{A. Pollo}
\author[b]{N. Produit}
\author[c]{D. Rybka}
\author[a]{J. Stauffer}
\author[a]{X. Wu}
\author[c]{A. Zadrozny}
\author[e,f]{S.N. Zhang}
\author[e]{J. Sun}
\author[e]{B. Wu}

\affil[a]{University of Geneva, 24 Quai Ernest-Ansermet, Geneva, Switzerland}
\affil[b]{University of Geneva, Geneva Observatory, ISDC, 16, Chemin d’Ecogia, 1290 Versoix Switzerland}
\affil[c]{National Centre for Nuclear Research, ul. A. Soltana 7, 05-400 Otwock, Swierk, Poland}
\affil[d]{Max-Planck-Institut fur extraterrestrische Physik, Giessenbachstrasse 1, D-85748 Garching, Germany}
\affil[e]{Key Laboratory of Particle Astrophysics, Institute of High Energy Physics, Chinese Academy of Sciences, Beijing 100049, China}
\affil[f]{University of Chinese Academy of Sciences, Beijing 100049, China}

\authorinfo{Further author information: (Send correspondence to J. Hulsman) \\
J. Hulsman : E-mail: johannes.hulsman@unige.ch, Telephone: +41 22 379 63 89 \\
POLAR-2 Collaboration: \url{https://www.astro.unige.ch/polar-2}
}

\begin{document}

\includepdf[pages=-]{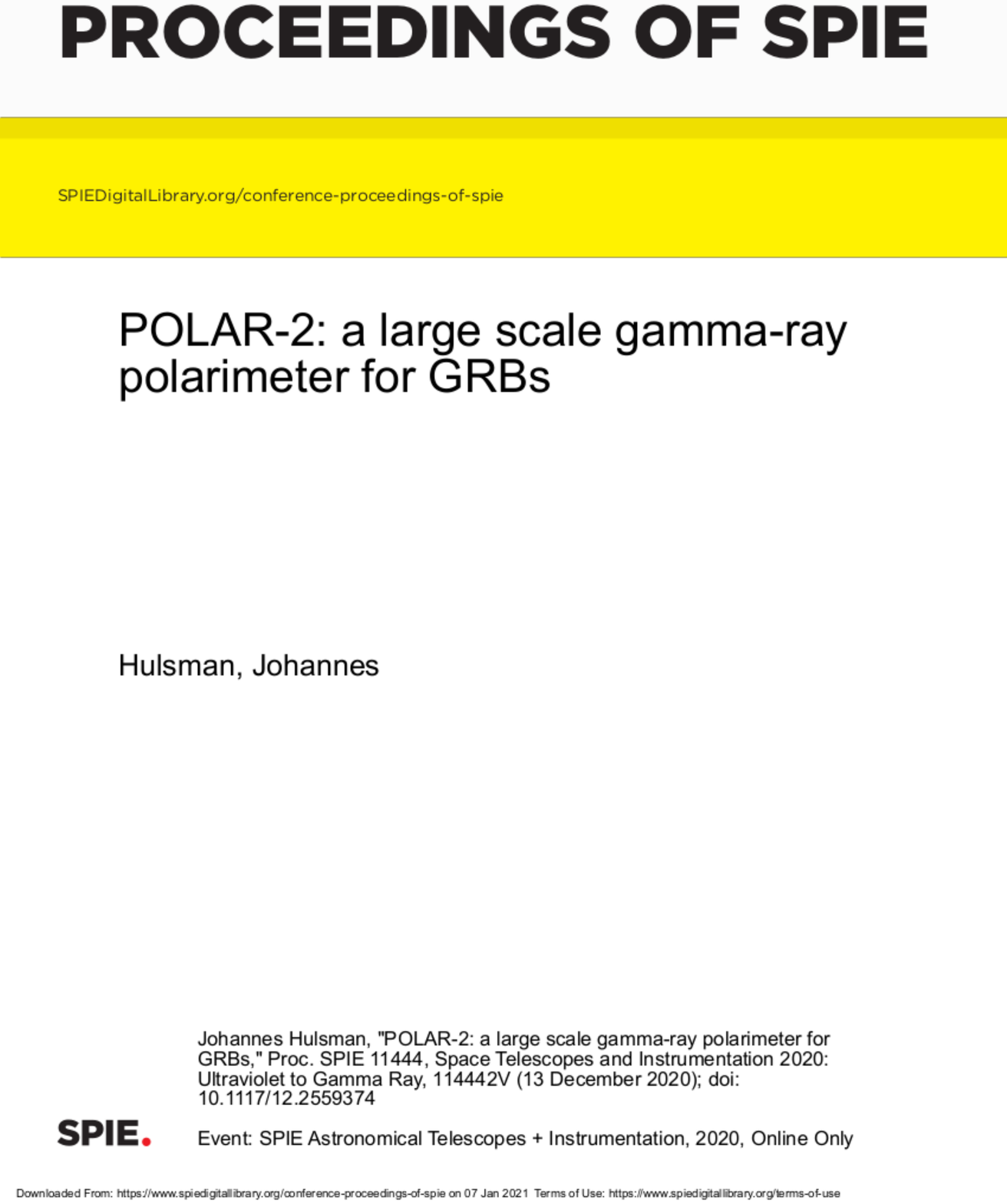}

\maketitle

\begin{abstract}

The prompt emission of GRBs has been investigated for more than $\SI{50}{years}$ but remains poorly understood. Commonly, spectral and temporal profiles of $\gamma$-ray emission are analysed. However, they are insufficient for a complete picture on GRB-related physics. The addition of polarization measurements provides invaluable information towards the understanding of these astrophysical sources. In recent years, dedicated polarimeters, such as POLAR and GAP, were built. The former of which observed low levels of polarization as well as a temporal evolution of the polarization angle. It was understood that a larger sample of GRB polarization measurements and time resolved studies are necessary to constrain theoretical models.
The POLAR-2 mission aims to address this by increasing the effective area by an order of magnitude compared to POLAR. POLAR-2 is manifested for launch on board the China Space Station in 2024 and will operate for at least 2 years. Insight from POLAR will aid in the improvement of the overall POLAR-2 design. Major improvements (compared to POLAR) will include the replacement of multi-anode PMTs (MAPMTs) with SiPMs, increase in sensitive volume and further technological upgrades. POLAR-2 is projected to measure about 50 GRBs per year with equal or better quality compared to the best seen by POLAR.
The instrument design, preliminary results and anticipated scientific potential of this mission will be discussed.

\end{abstract}

% Include a list of keywords after the abstract 
\\
\\
\newpage
\section{INTRODUCTION}
\label{sec:intro}

Gamma-ray bursts (GRBs) are the brightest and most energetic electromagnetic astrophysical processes in the known Universe with an emitted energy of up to $\SI{E53}{erg}$. Early measurements by the Burst and Transient Experiment (BATSE) \cite{batse_1992} found these events to be isotropically distributed and hence extragalactic. BATSE also showed that GRBs could be divided into two primary subclasses: long and short bursts with for t $>\SI{2}{s}$ and t $<\SI{2}{s}$ respectively. Either prompt emissions are followed by a longer lived "afterglow", emitting wavelengths from radio waves to TeV energies. Its first detection was during the measurement of the long GRB \emph{970228} by Beppo-SAX \cite{Beppo_sax_92}. LIGO, VIRGO, INTEGRAL and Fermi-GBM have shown through a joint analysis of \emph{170817A} that short GRB burst originate from binary neutron star mergers \cite{Abbott_2017}. Growing evidence indicates that long GRBs are accompanied by massive star explosions (i.e. GRB-SN events) \cite{Woosley_2006}. These events are more energetic and luminous than conventional SNs events. Current GRB-SN events suggest that most of the energy is contained in the nonrelativistic ejecta (producing the SN) rather than in the relativistic jets producing the GRB and its afterglow \cite{Woosley_2006}.
\newline

Despite extensive analysis on the spectral and temporal profile of $\gamma$-ray emissions of astrophysical sources, several key questions regarding the physics of the central engines of relativistic jets and the jets themselves remain unresolved or sometimes even impossible to answer \cite{toma2013polarization}. The addition of polarization information addresses some of these ambiguities. In particular, the polarization degree (PD) and polarization angle (PA) can illuminate our understanding on the emission mechanism, geomagnetic structure of GRB jets and magnetic composition. Theoretical models describing synchrotron radiation from large-scale dynamic magnetic fields present linear PDs up to $56\%$ \cite{Lyutikov_2003}. If those magnetic fields are highly ordered the average PD can be $40\%$ for a sample of GRBs \cite{toma2013polarization}. In \cite{Lundman_2018}, the MeV spectral peak of GRBs is best described through a photospheric emissions model. There, linear PDs are of the order of a few percent above $\SI{100}{keV}$, increasing to about $50\%$ below $\SI{1}{keV}$. However, the model also predicts an increase of PDs (up to $40\%$) for jets seen at larger off-axis angles \cite{Lundman_2018}. \newline

To properly distinguish between these differing models, one would need a large sample of GRB measurements which are seen by a detector system capable of measuring the temporal evolution of the linear polarization parameters on an event by event basis. These challenges were initially addressed by POLAR and its successor POLAR-2; the latter of which is currently in development for a launch in 2024. \newline

\section{Gamma-Ray Burst Polarimetry Measurements}
\label{sec:grb_measurements}

The scientific benefits provided through polarization measurements of $\gamma$-rays has motivated a series of polarimetry instruments to probe the astrophysical sources. Several successful missions include GAP ($\SI{50}{keV}$ to $\SI{300}{keV}$) \cite{Yonetoku_2011}, COSI ($\SI{200}{keV}$ to $\SI{5}{MeV}$) \cite{Lowell_2017}, PoGOLite Pathfinder ($\SI{25}{keV}$ to $\SI{240}{keV}$) \cite{Chauvin_2015} and more recently POLAR ($\SI{50}{keV}$ to $\SI{500}{keV}$) \cite{kole2020polar}. Future polarimeter instruments proposed or under construction include the PRAXyS ($\SI{2}{keV}$ to $\SI{10}{keV}$) \cite{Iwakiri_2016}, XL-Calibur ($\SI{15}{keV}$ to $\SI{80}{keV}$) \cite{Abarr_2021}, LEAP ($\SI{30}{keV}$ to $\SI{500}{keV}$) \cite{LEAP_proc}, eXTP ($\SI{0.5}{keV}$ to $\SI{30}{keV}$) \cite{Zhang_2016}, IXPE ($\SI{2}{keV}$ to $\SI{8}{keV}$) \cite{Ferrazzoli_2020} and POLAR-2 ($\SI{20}{keV}$ to $\SI{800}{keV}$); the latter of which is the successor to POLAR. \newline

\begin{figure}[b]
  \centering
  \begin{subfigure}[b]{0.56\textwidth}
     \centering
     \includegraphics[width=\textwidth]{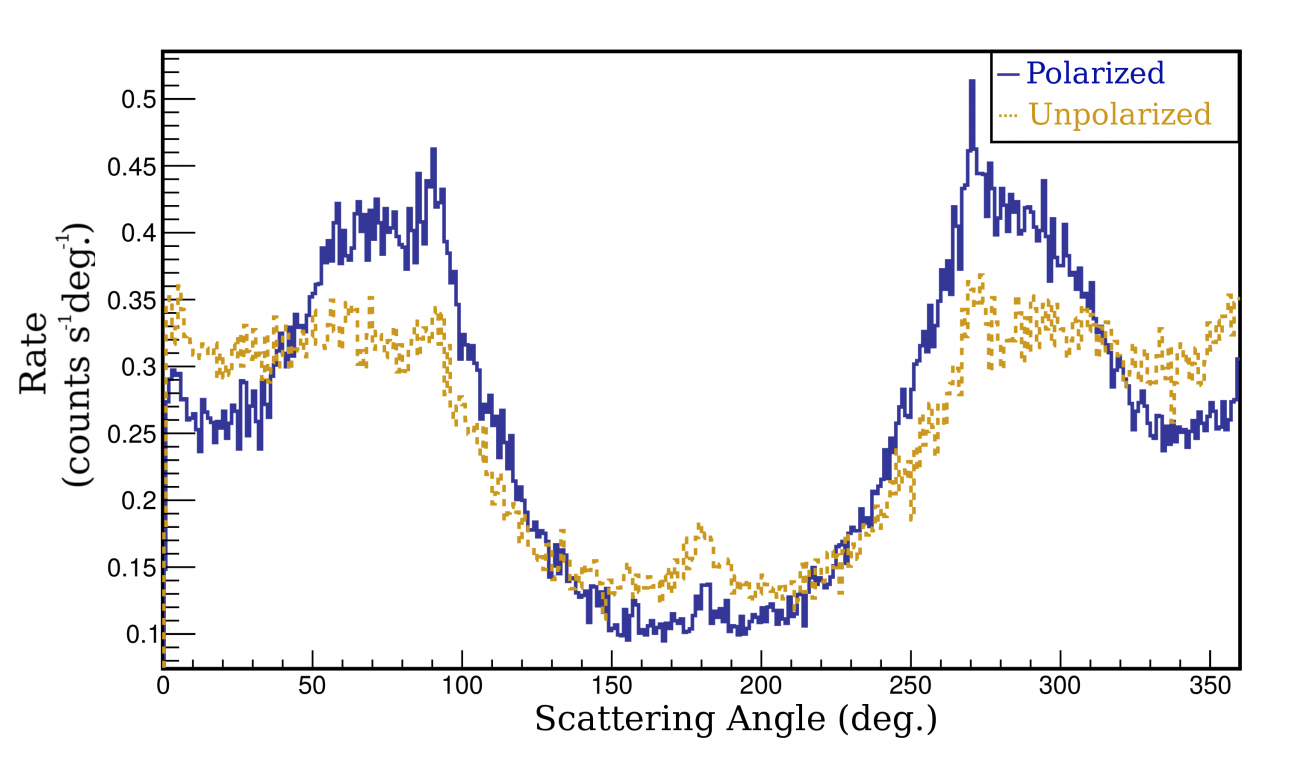}
     \caption{}
     \label{fig:polar_modulation}
  \end{subfigure}
  \hfill
  \begin{subfigure}[b]{0.4\textwidth}
     \centering
     \includegraphics[width=\textwidth]{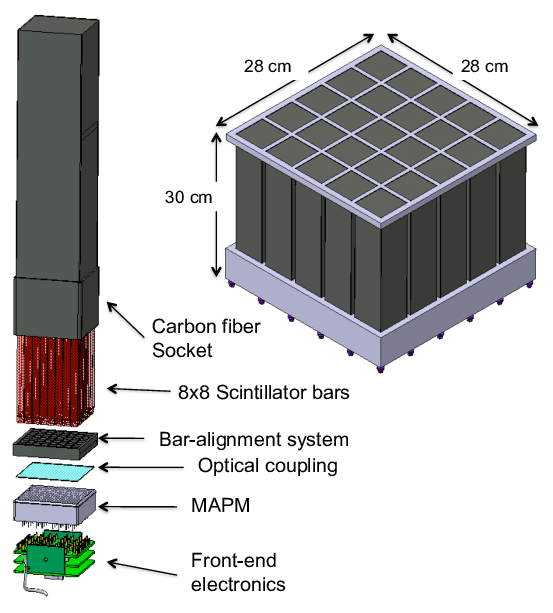}
     \caption{}
     \label{fig:polar_scheme}
  \end{subfigure}
  \caption{\textbf{a)} Simulated modulation curve of a polarized (blue) and unpolarized (yellow) scattering angle distribution (including instrument induced effects) \cite{kole2020polar} \textbf{b)} Schematic of POLAR instrument and one of its 25 modules \cite{kole2016polar}}
  \hfill
\end{figure}

%\textcolor{red}{add GAP results???}
To date, the most extensive and detailed analyses on GRB polarizations are obtained through POLAR. Its instrument design and analysis of 55 GRBs are outlined in \cite{Produit_2018} and \cite{kole2020polar} respectively. POLAR was a dedicated polarimeter which was selected to be launched in 2016 to the Chinese space laboratory Tiangong-2 (TG-2). It was comprised of 25 modules (in a 5x5 layout), each housing 64 thin and elongated plastic scintillator bars (in a 8x8 layout) with dimensions $5.8\times 5.8\times 176\SI{}{mm}$. Scintillators inside the module are then read out with their respective 64 channel Multi-Anode PMT (MAPMT). Each module also contains its respective front-end electronics (FEE). A schematic of the POLAR instrument and one of its modules is shown in figure \ref{fig:polar_scheme}. Combined, it provides an effective area of $\SI{400}{cm^2}$ at $\SI{350}{keV}$. Polarization information is obtained from photons which scatter between two scintillators within $\SI{100}{ns}$, the second interaction being through Compton scattering or photo-absorption. This process can be described by the Klein-Nishina formula which includes the polar scattering angle and azimuthal scattering angle to the polarization vector \cite{kole2016polar}. The final distribution of the azimuthal scattering angle has a sinusoidal structure which is commonly referred to as the \emph{modulation curve}. An example of such a modulation curve is shown in \ref{fig:polar_modulation} for a polarized and unpolarized photon flux (including detector induced effects). As can be seen, the distribution shows a clear $\SI{180}{^\circ}$ modulation and a smaller non-negligible $\SI{360}{^\circ}$ modulation from the induced polarized flux. \newline

\begin{figure}[t]
  \centering
  \begin{subfigure}[b]{0.49\textwidth}
     \centering
     \includegraphics[width=\textwidth]{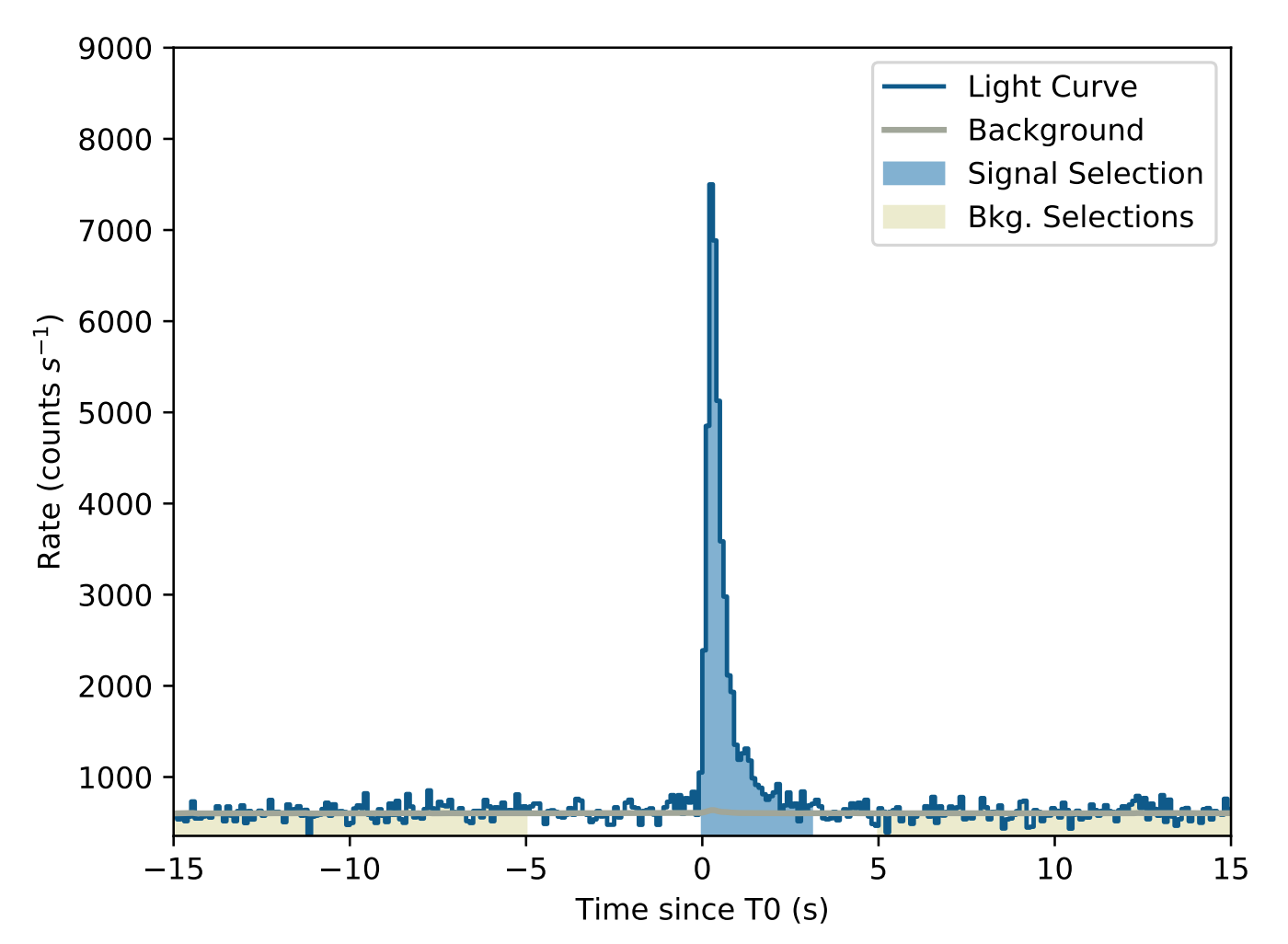}
     \caption{}
     \label{fig:grb_lightcurve}
  \end{subfigure}
  \hfill
  \begin{subfigure}[b]{0.49\textwidth}
     \centering
     \includegraphics[width=\textwidth]{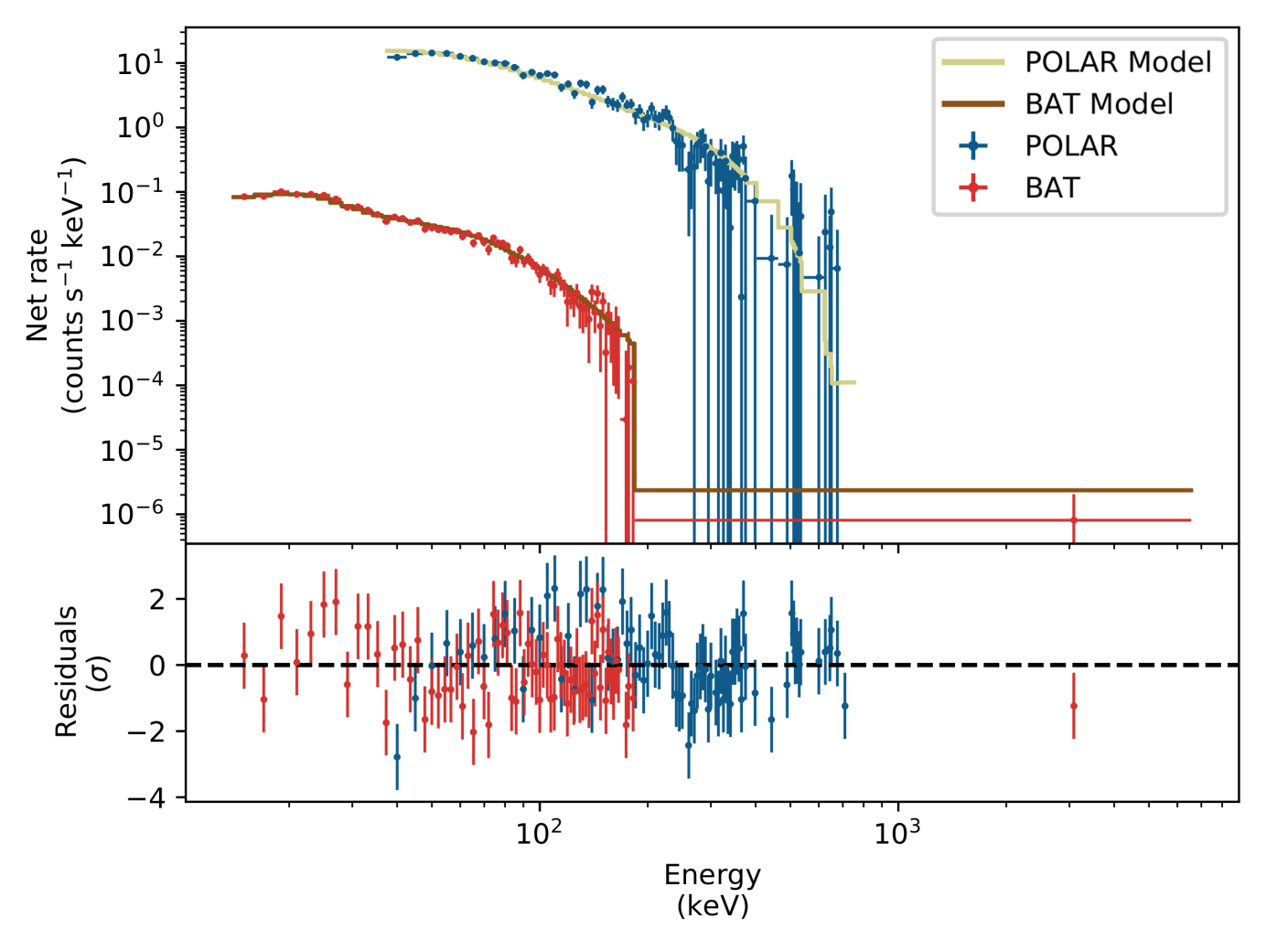}
     \caption{}
     \label{fig:grb_spectrum}
  \end{subfigure}
  \caption{\textbf{a)} Light curve of GRB170101A measured by POLAR. For this GRB, T0 is defined by \emph{Swift}-BAT. \textbf{b)} A combined spectral fit as measured by POLAR (blue) and \emph{Swift}-BAT (red). The fitted energy spectrum, folded with the detector response is shown in yellow (POLAR) and brown (\emph{Swift}-BAT). \cite{kole2020polar}}
  \hfill
\end{figure}

After its launch in September 2016, the POLAR instrument measured 55 GRBs until a high voltage power supply failure ended its campaign after 6 months \cite{kole2020polar}. A sample GRB (GRB170101A) is shown in figures \ref{fig:grb_lightcurve} (light curve) and \ref{fig:grb_spectrum} (energy spectrum folded with the detector response). It should be noted that the  mission proved so successful that during its course the data bandwidth on the space station was increased. 
Based on these GRBs it was found that from time integrated results, prompt emissions between $\SI{30}{keV}$ and $\SI{750}{keV}$ are unpolarized (or are very lowly polarized). This applies to long and short GRBs as well as GRBs with single or multiple peaks. This is in disagreement with results from  AstroSAT \cite{Chattopadhyay_2019} where polarization degrees are found to be above 50\%. A future joint analysis should address this disagreement. Further POLAR results exclude theoretical models of a synchrotron emission with an ordered toroidal magnetic field \cite{kole2020polar}. They are however compatible with Compton drag models and synchrotron emissions models with a radial magnetic field. Results are also consistent with photospheric emission models described in \cite{Lundman_2018} which are generally around 0\% for the energy range in question (up to 40\% for larger off-axis angles). The temporal evolution of the PD and PA seen by POLAR are in agreement with most theoretical model predictions. No significant changes were found in the PD of multi-peak GRBs \cite{kole2020polar}. However, trace results hint towards a possible PA evolution within single peak GRBs. No known physical model fully explains this and further time resolved theoretical modelling is required. \newline
Despite the vast majority of information gathered from POLAR data, more could be extracted. Energy resolved studies can help probe the photospheric emissions model by \cite{Lundman_2018} which shows non-zero polarization degrees at low energies (10's of keV). Nevertheless, no analysis can resolve the need for higher precision measurements. In combination with longer mission, larger GRB catalogues could be produced with more detailed temporal and energy resolved analysis; thus helping our understanding of the physics of GRBs. Experiences from the POLAR instrument are carried into the development of the POLAR-2 instrument which is manifested for launch on board the China Space Station in 2024. \newline

\section{Development of the POLAR-2 Instrument}
\label{sec:polar2_instrument}

\subsection{POLAR-2 Instrument Design}
\label{sec:polar2_design}

As discussed in chapter \ref{sec:grb_measurements}, the polarization within a GRB is obtained from the scattering angle of photons between at least two scintillator bars within a time window of $\SI{100}{ns}$; events triggering only single bars are therefore mostly discarded. We therefore adopt a similar modular design as for POLAR, where each module will house 64 scintillators, their respective photomultiplier and FEE. Here, each scintillator will be $5.9\times5.9\times125\SI{}{mm}$ and wrapped with Claryl foil \cite{clary_datasheet} to minimize photon loss (produced through scintillation) as well as preventing it from entering its neighbouring scintillator. Furthermore, these scintillators will not be truncated as in the POLAR design (which were necessary to reduce the cross talk) and thus increasing the scintillator area to the photo-multiplier as well as reducing photon loss through total internal reflection and conservation of phase space (Liouville's Theorem). These scintillators will also be composed of EJ-200 (as opposed to EJ-248M), yielding a higher scintillation efficiency (9\% more) and a longer light attenuation length (52\% longer). 
Contrary to POLAR, these scintillators will be read out by Hamamatsu SiPMs (as opposed to MAPMTs). This change provides three advantages as it i) simplifies the mechanical design (SiPM are more robust compared to MAPMTs and thus do not require dead material for damping purposes), ii) eliminates the need to develop a space qualified high voltage power supply (which was the only component to fail during the POLAR mission) and iii) allows us to lower the low energy detection threshold. It will be glued with a silicon based adhesive (MAPSIL QS1123 \cite{Guillaumon2009DEVELOPMENTOA}) which also serves as an optical coupling. For POLAR, no better than 0.3 photoelectrons/keV could be obtained. We aim to increase this to possibly 1.5 photoelectrons/keV. The net result will be a higher effective area at lower energies, thus allowing us to test theoretical models such as \cite{Lundman_2018}. However, the drawback of SiPMs are their increased dark noise at room temperatures. Operating POLAR-2 at sub-zero temperatures would reduce this issue. The addition of a Peltier element can cool the SiPM below $\SI{-10}{^\circ C}$, provided that the heat can be efficiently extracted. In addition to the polarimeter, POLAR-2 will also include multiple spectrometers, inspired by GECAM which reports an energy resolution of $<8\%$ between $\SI{6}{keV}$ and $\SI{5}{MeV}$ \cite{Zhang_2019}. \newline

\begin{figure}[t]
  \centering
  \begin{subfigure}[b]{0.49\textwidth}
     \centering
     \includegraphics[width=0.8\textwidth]{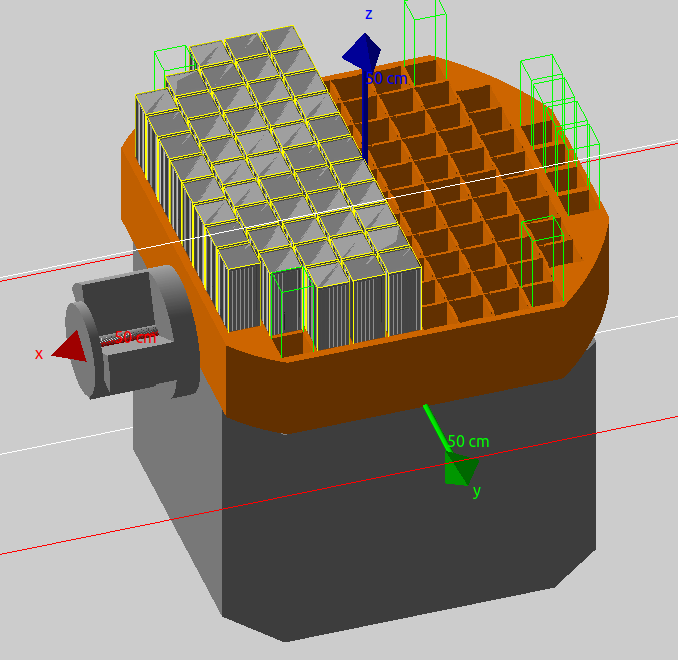}
     \caption{}
     \label{fig:polar2_geant4}
  \end{subfigure}
  \hfill
  \begin{subfigure}[b]{0.49\textwidth}
     \centering
     \includegraphics[width=0.33\textwidth]{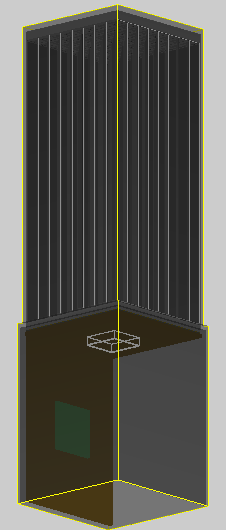}
     \caption{}
     \label{fig:Module_carbon}
  \end{subfigure}
  \caption{\textbf{a)} Mass model of the POLAR-2 instrument as it is currently implemented for simulation purposes (within the Geant4 framework). \textbf{b)} A close-up of the POLAR-2 polarimeter module inside its carbon fiber casing. The green square reflects a potential FPGA, whereas the white/transparent object beneath the scintillators marks the Peltier device.}
  \label{fig:polar2_sim}
  \hfill
\end{figure}

In figures \ref{fig:polar2_geant4} and \ref{fig:Module_carbon} we illustrate the design of the POLAR-2 module as it is implemented with the Geant4 simulation framework \cite{AGOSTINELLI2003250} (enabling detailed particle physics simulations). Figure \ref{fig:Module_carbon} shows the preliminary design of a POLAR-2 module inside its carbon fiber frame. The top encapsulates the scintillators, whereas the bottom houses the mechanical support, FEE and thermal cooling system. These modules are then placed inside an aluminium grid (colored orange in figure \ref{fig:polar2_geant4}) such that only the scintillators face the sky. The grid provides mechanical support but also shields the electronics from background radiation, most importantly those produced from solar flares or by passing through the South Atlantic Anomaly (SAA). 100 modules will be placed inside the grid, already quadrupling what was done for POLAR. 8 empty sockets marked with green rectangular volume edge lines highlight where spectrometers could be potentially placed (studies of the optimal spectrometer location are ongoing). \newline
Below the aluminium grid is a hollow cube-like base with an adapter (left in figure \ref{fig:polar2_geant4}). Its primary purpose is mechanical support during launch and placement on the China Space Station. The mechanical support and polarimeter module will have a combined dimension of about $650\times650\times665\SI{}{mm}$ with each module being about $55\times55\times200\SI{}{mm}$. \newline

\begin{figure}[t]
  \centering
  \begin{subfigure}[b]{0.49\textwidth}
     \centering
     \includegraphics[width=0.8\textwidth]{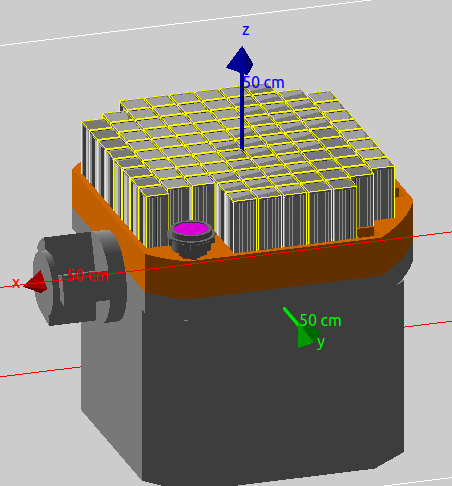}
     \caption{}
     \label{fig:polar2_gecam}
  \end{subfigure}
  \hfill
  \begin{subfigure}[b]{0.49\textwidth}
     \centering
     \includegraphics[width=1.03\textwidth]{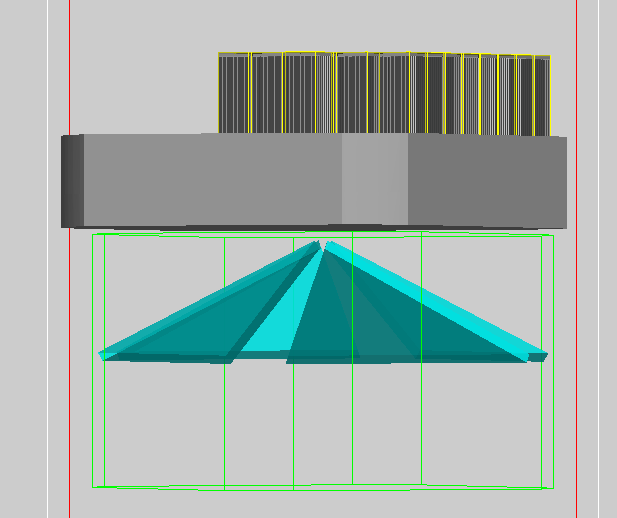}
     \caption{}
     \label{fig:spectrometer_pyramid}
  \end{subfigure}
  \caption{\textbf{a)} Placement of an exact GECAM replica placed at one of its designated sockets as marked in figure \ref{fig:polar2_geant4}. \textbf{b)} Alternative spectrometer design where the active volume has a pryamid-like design, placed inside the mechanical support structure of POLAR-2}
  \label{fig:polar2_spectr}
  \hfill
\end{figure}

Several GRBs seen by POLAR were not observed by $\gamma$-ray spectrometers or for which the localization measurement was outside the $\SI{1}{^\circ}$ level. The lack of these complimentary measurements resulted in larger systematic errors within the polarization measurements and should thus be avoided. This can be achieved by including spectrometers as part of the POLAR-2 instrument oriented such that their field of view covers half the sky. Potential designs and placements of spectrometers are illustrated in figure \ref{fig:polar2_spectr}. In figure \ref{fig:polar2_gecam} we see the placement of a GECAM replica as characterized in \cite{Zhang_2019}. The active volume is a LaBr$_3$:Ce scintillator crystal which is generally better compared to conventional crystals such as NaI or CsI as it features a higher light yield and energy resolution. Furthermore, its light output remains stable during temperature fluctuations experience in space. However, background radiation activates the crystal, increasing its internal background (as well as the background seen by the polarimeter) and thus decreases its overall performance with increasing orbital times. Eventually, the activation could obstruct the performance of the polarimeter. Feasibility analyses (accounting for the science case) are ongoing to study this effect. We also see in figure \ref{fig:polar2_gecam} that the current GECAM design is relatively bulky and reduces number of polarimeter modules on the aluminium grid. As a result, to not interfere with the polarimeter, the GECAM could be reduced in size or placed elsewhere. Simulations are currently performed to study the optimal position and size. Alternatively, a pyramid-like structure (shown in figure \ref{fig:spectrometer_pyramid}) could be placed below the polarimeter inside the mechanical support structure (which is hollow). Its active volume is CeBr$_3$ which has a comparatively lower energy resolution compared to LaBr$_3$ \cite{crystal_compare}. However, it has almost no naturally occurring radioactive isotopes, resulting in a significantly lower internal background. The configuration shown in figure \ref{fig:spectrometer_pyramid} (intended for a case study) has a total surface area of about $\SI{2600}{cm^2}$ ($4\times\SI{650}{cm^2}$). Each triangle is $\SI{410}{mm}$ wide at its base, $\SI{310}{mm}$ high and $\SI{15}{mm}$ thick. They are placed such that they have an opening angle of about $\SI{121}{^\circ}$. Simulations indicate (for the design in figure \ref{fig:spectrometer_pyramid}) that the effective area for $\SI{1}{MeV}$ vertically impinging $\gamma$'s (wrt. the z-axis indicated in figure \ref{fig:polar2_gecam}) to be about $\approx \SI{500}{cm^2}$. \newline

\subsection{POLAR-2 Module Performance}
\label{sec:polar2_performance}

In this section we will discuss some of the work and results performed towards the overall development of the POLAR-2 module including some initial tests performed with a CeBr3 crystal for spectrometer purposes. Studies regarding the space qualification of commercial-off-the-shelf (COTS) parts, thermal cooling and choice of reflective foil will be discussed. In \ref{module_prototype} preliminary results of a POLAR-2 module are shown. \newline

\subsubsection{Space Qualification and Radiation Tests}

As POLAR-2 is a space-borne detector it must withstand the radiation in low earth orbit. Electronics are subjected to irradiation from cosmic particles (mostly from the Sun). Different particles deposit the energy differently, yielding differing doses. Higher doses in the silicon die of an FPGA (in the FEE) results in eg. bit flips, which in turn causes a faulty performance. In space, these can be averted by power cycling the board. Nevertheless, it is pertinent to understand the frequency of such a procedure. Preferentially, the time interval between each power cycle should be as long as possible. An FPGA which requires, for e.g., a power cycle every hour would be unsuitable/unreliable for space operation. \newline

\begin{figure}[t]
  \centering
  \begin{subfigure}[b]{0.49\textwidth}
     \centering
     \includegraphics[width=0.6\textwidth]{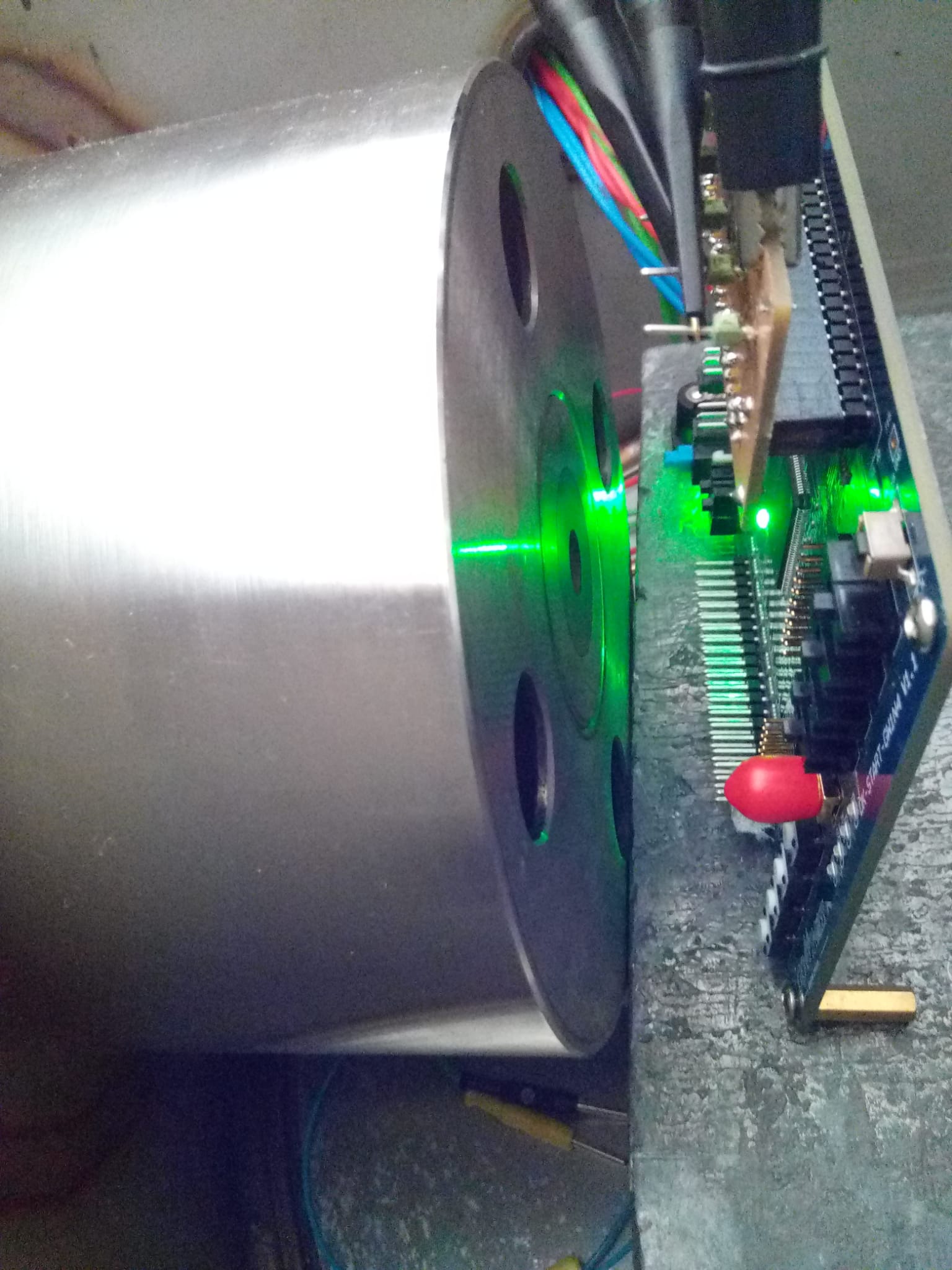}
     \caption{}
     \label{fig:gowin_dpnc}
  \end{subfigure}
  \hfill
  \begin{subfigure}[b]{0.49\textwidth}
     \centering
     \includegraphics[width=0.6\textwidth]{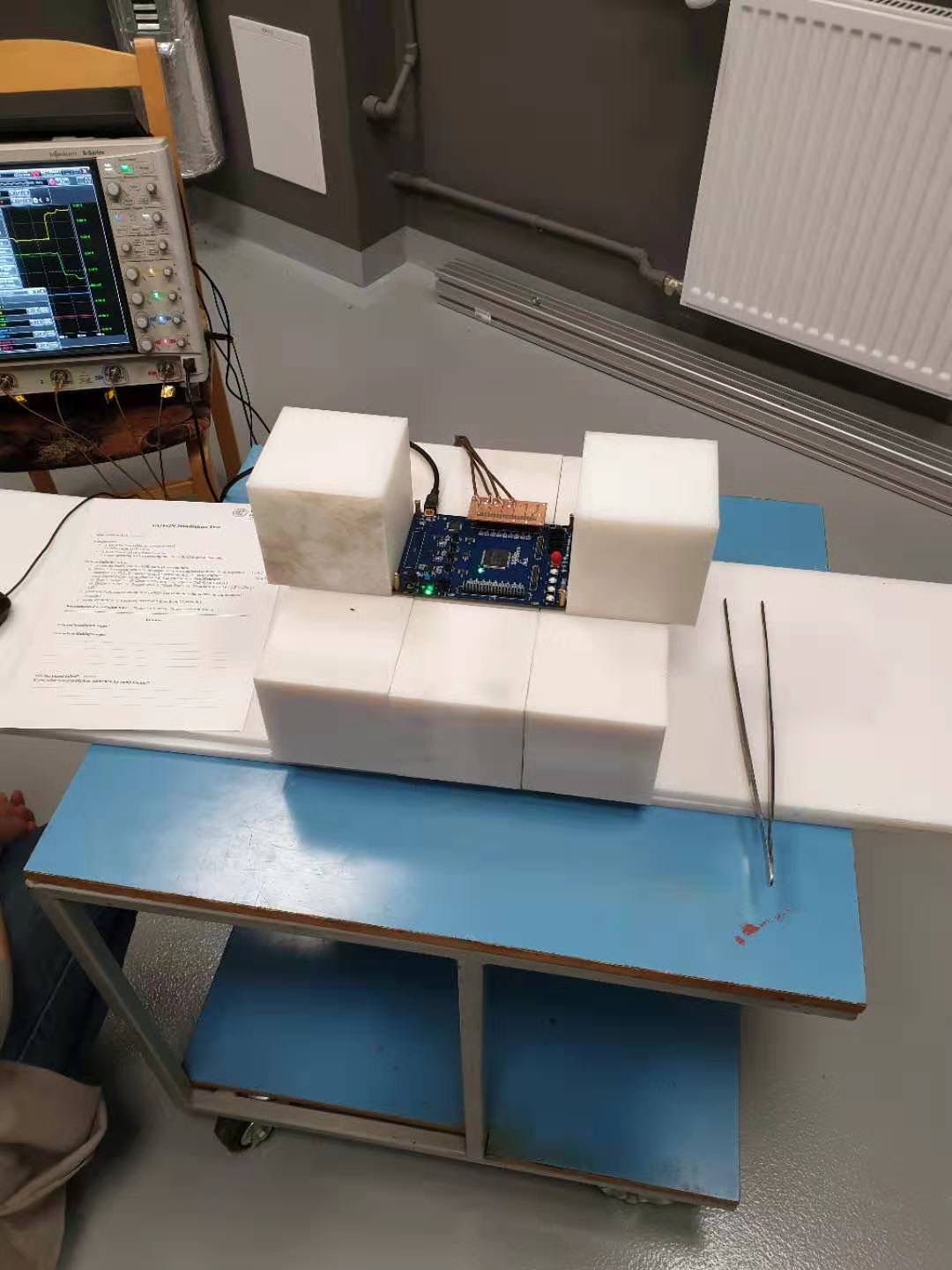}
     \caption{}
     \label{fig:gowin_ncbj}
  \end{subfigure}
  \caption{GOWIN development board set for its irradiation with a \textbf{a)} $^{90}$Sr and \textbf{b)} PuBe neutron source.}
  \label{fig:gowin_irradiate}
  \hfill
\end{figure}

The operation of POLAR in space has shown that an IGLOO FPGA \cite{Produit_2018} is reliable for space-borne missions. In our pursuit for possibly better and cheaper FPGAs alternatives we also looked into the FPGA provided by \emph{Gowin Semiconductor Corp.}. A radiation test was developed to investigate the effects of the total ionizing dose (TID), transient dose effect (TDE) and single event effects (SEE). The TID is the damage due to increasing exposure to radiation. The TDE described the effect of transistors to open/close due to short-time radiation pulses. The SEE describes bit flips in the memory due to the ionized track left by the particles. The architecture is based on a chain ring shift register which alternates between 0 and 1 (this was flashed on the GOWIN FPGA). Any bit flip would switch the 0/1 into 1/0 and flag as an error (as you would have e.g. '\color{green}0\color{red}0\color{green}01\color{black}' instead of '\color{green}0101\color{black}'). The FPGA is then irradiated with a $^{90}$Sr (figure \ref{fig:gowin_dpnc}) and a PuBe (figure \ref{fig:gowin_ncbj}) source; the latter of which emitting neutrons and $\gamma$'s up to $\SI{10}{MeV}$. In combination with dedicated simulations, it was found that the cumulative dose through $^{90}$Sr and PuBe equates to 13 years of operation in space\footnote{at an altitude of $\SI{300}{km}$ and an inclination of $\SI{42}{^\circ}$}\footnote{600 years for a more conservative approach through the addition of shielding from the CSS}. During the course of its irradiation no bit flips were encountered, indicating that it could be used for our mission. \newline

\begin{figure}[t]
  \centering
  \begin{subfigure}[b]{0.49\textwidth}
     \centering
     \includegraphics[width=1\textwidth]{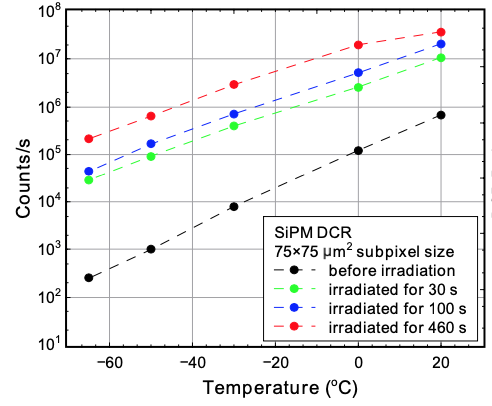}
     \caption{}
     \label{fig:sipm_dcr}
  \end{subfigure}
  \hfill
  \begin{subfigure}[b]{0.49\textwidth}
     \centering
     \includegraphics[width=1\textwidth]{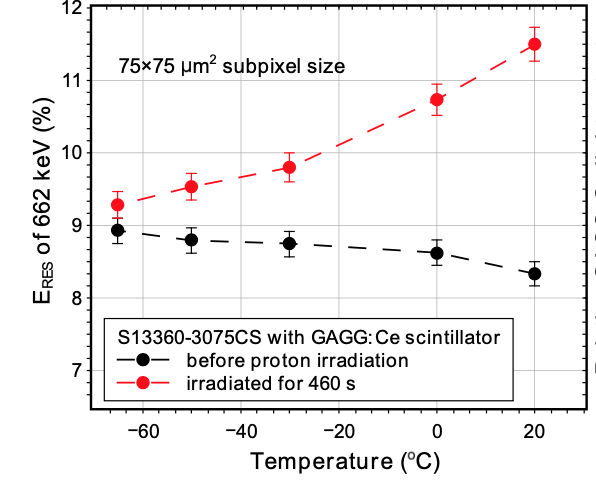}
     \caption{}
     \label{fig:sipm_resolution}
  \end{subfigure}
  \caption{\textbf{a)} Dark current for a $75\times75 \SI{}{\mu m^2}$ microcell pixel for different proton (\SI{170}{MeV}) fluences. \textbf{b)} Energy resolution from a $\SI{662}{keV}$ line ($^{137}$Cs source) before and after irradiating the subpixel. \cite{Mianowski_2020}}
  \label{fig:sipm_paper}
  \hfill
\end{figure}

SiPMs are also susceptible to the background radiation. This was already demonstrated and studied for the Strontium Iodide Radiation Instrument (SIRI) \cite{mitchell2020radiation} where the SiPM bias current has increased by $\SI{528}{\mu A}$ over the course of its 1-year mission (typical dark current at $\SI{28.9}{V}$ is about $\SI{10}{\mu A}$). Note, for SIRI, 2x2 J-series 60035 SiPM arrays were used. However, it was found that SensL SiPMs perform similarly to the S13360-6050CS (similar model to the one intended to be used for POLAR-2; S13361-6075NE-04). Another similar SiPM, the S13360-3075CS, had been irradiated with $\SI{170}{MeV}$ protons with an average flux of $\SI{e7}{\frac{protons}{s\times cm^2}}$ and discussed in \cite{Mianowski_2020}. There, the channel size is smaller. However, the microcell size is equivalent to the one we consider to use (and thus yield comparable results). As shown in figure \ref{fig:sipm_dcr}, we see the dark current decrease for lower temperatures. Furthermore, as expected, the dark current increases with increasing radiation times. Assuming the Peltier device will cool the SiPM pixels to $\SI{-20}{^\circ C}$, we expect its usability in space to be of the order of $\SI{e8}{s}$ (i.e. several years) \cite{Mianowski_2020} with its energy resolution decreasing from 8.7\% to 10.2\% (obtained from a $\SI{662}{keV}$ line
from a $^{137}$Cs source). Equivalent analysis on the S13361-6075NE-04 are ongoing. \newline

The Peltier primarily serves to cool the SiPM pixels to $\SI{-20}{^\circ C}$. Due to its proximity to the SiPM, brief studies were performed on a Peltier of type PE-031-10-15-S by irradiating it with the PuBe neutron source. Solar flare events or SAA passings should not activate the material as this would subsequently increase the number of background events. The neutron source has flux of about $\SI{7.5e5}{\frac{neutrons}{4\pi\times s}}$ and was placed directly on top of the Peltier for 12 hours. No isotopes with long decay times and whose decay products are in the keV or MeV range were measured. This is in agreement with simulations were most isotopes were found to be in the $\SI{e-15}{s}$ range\footnote{Note, the instrument will be powered down or stop data acquisition while passing through the SAA}. This is beneficial for our mission. \newline

\subsubsection{Reflective Foil}
\label{refl_foil}

Reflectance and transmittance measurements had been performed on Claryl reflective foils\footnote{from Toray} to find suitable thinner alternatives to Vikuiti (3M VM2000 ESR) which was used for POLAR. They are primarily made of plastic, coated with a thin reflective aluminium layer on one side (the reflective layer facing the scintillator bar surfaces). Thinner foils allow scintillators to be thicker (increase in sensitive volume). An alternative foil should potentially increase the light yield within a bar. \newline

\begin{figure}[t]
  \centering
  \begin{subfigure}[b]{0.385\textwidth}
     \centering
     \includegraphics[width=\textwidth]{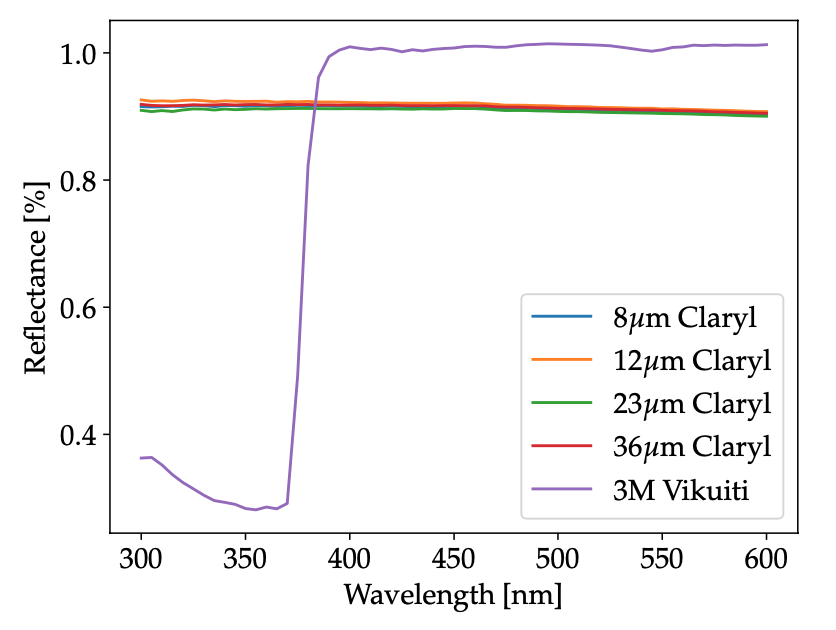}
     \caption{}
     \label{fig:claryl_refl}
  \end{subfigure}
  \hfill
  \begin{subfigure}[b]{0.605\textwidth}
     \centering
     \includegraphics[width=\textwidth]{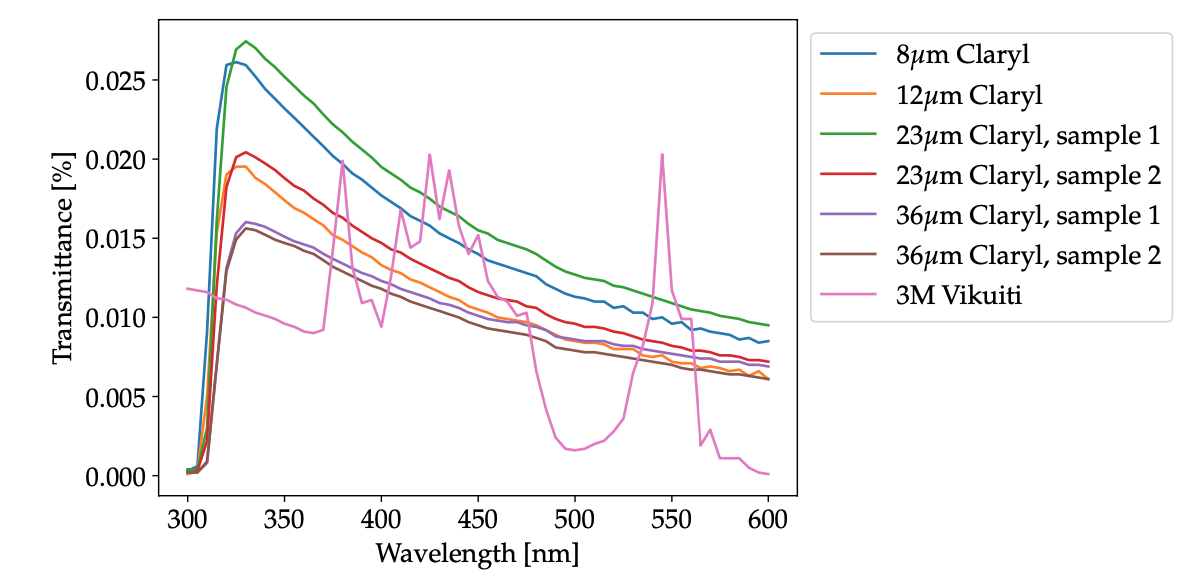}
     \caption{}
     \label{fig:claryl_trans}
  \end{subfigure}
  \caption{\textbf{a)} Reflectance and \textbf{b)} transmittance spectrum shown for various Claryl thicknesses compared to a $\SI{65}{\mu m}$ Vikuiti film.}
  \label{fig:claryl_spectral}
  \hfill
\end{figure}

Figures \ref{fig:claryl_refl} and \ref{fig:claryl_trans} show the total reflectance and transmittance of differing Claryl foil thicknesses compared to a $\SI{65}{\mu m}$ Vikuiti (as was used for POLAR) respectively. We find that Vikuiti has a higher reflectance compared to Claryl (~100\% vs 85\%). Nevertheless, Claryl and Vikuiti have similar average transmittances (of the order 0.012\%). This results from the higher absorbance from the aluminium layer on the film. No significant differences are seen between differing Claryl film thicknesses. 
As the transmittance is of key importance (prevents photons to pass to adjacent scintillator bars), we find it to be a suitable replacement for Vikuiti. Note, Vikuiti foils were placed between scintillator bars, leaving small gaps at the scintillator corners. Wrapping these scintillators with Claryl foils prevents this from happening as well as effectively doubling the number of foils between scintillator bars. Accounting for practical reasons, a Claryl thickness of $\SI{12}{\mu m}$ was selected for the final design. The thinner film has allowed the sensitive volume to increase by 3.5\%.

\begin{figure}[t]
  \centering
  \begin{subfigure}[b]{0.42\textwidth}
     \centering
     \includegraphics[width=0.7\textwidth]{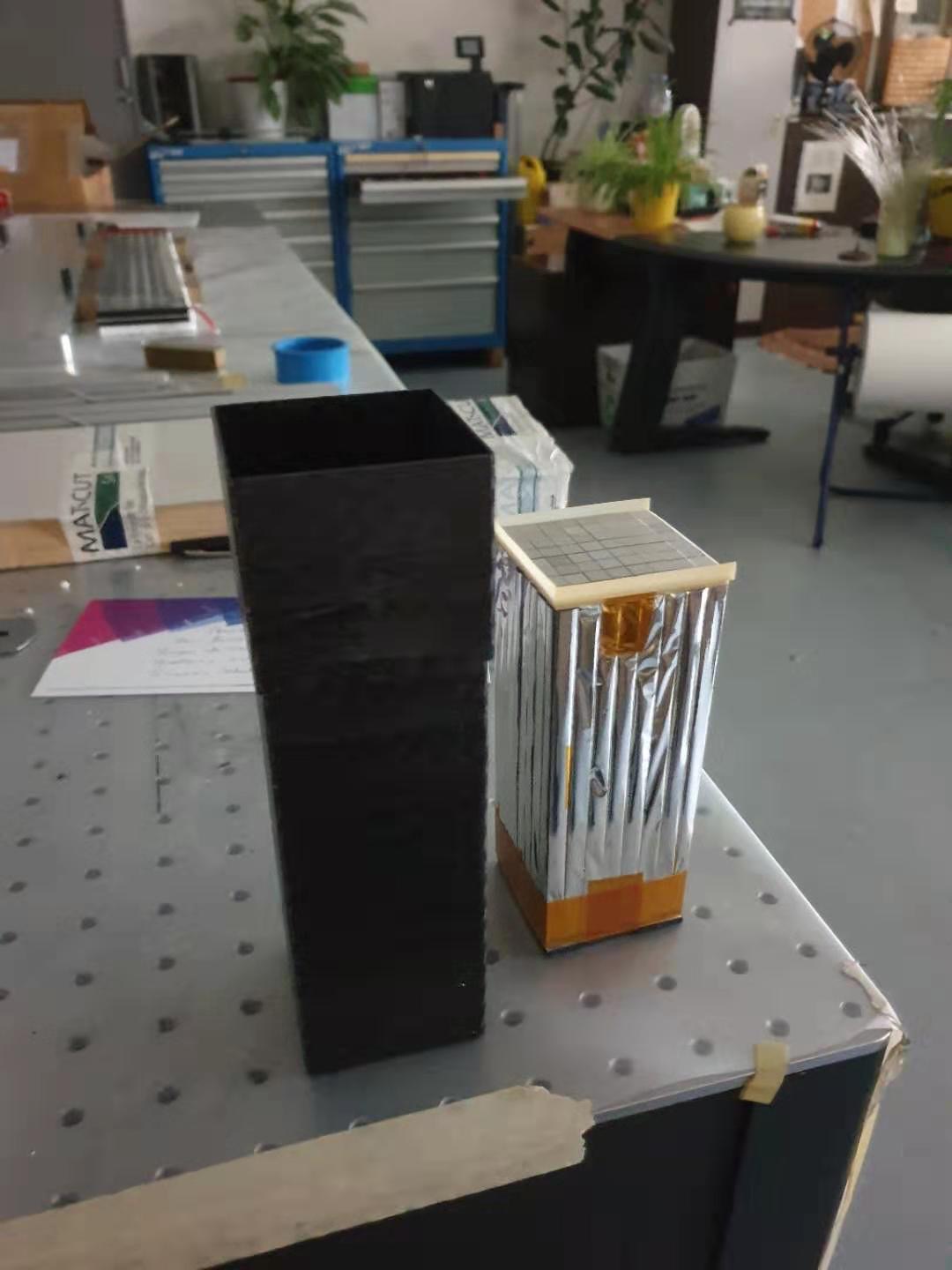}
     \caption{}
     \label{fig:module_exposed}
  \end{subfigure}
  \hfill
  \begin{subfigure}[b]{0.57\textwidth}
     \centering
     \includegraphics[width=0.7\textwidth]{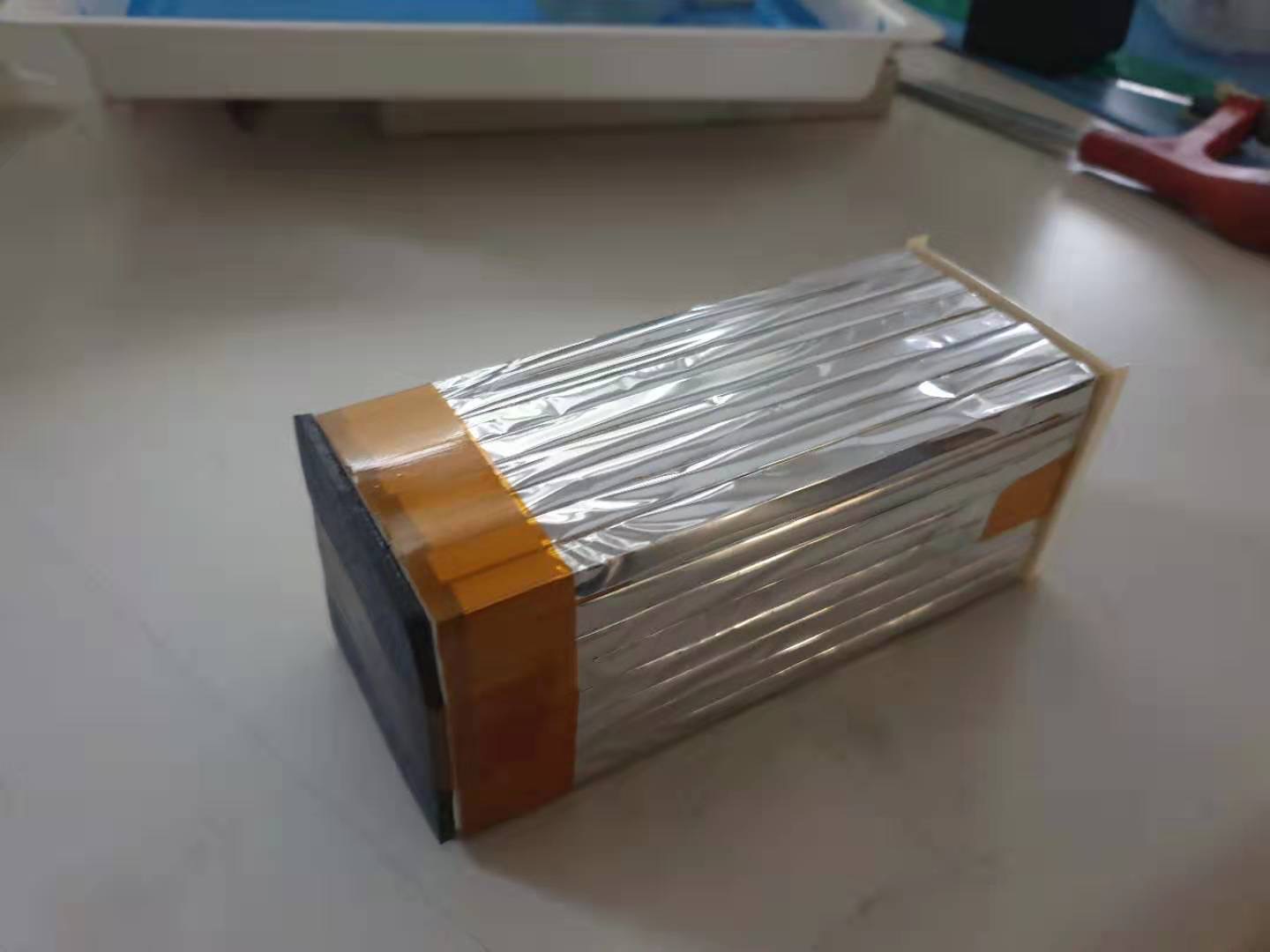}
     \caption{}
     \label{fig:claryl_proto2}
  \end{subfigure}
  \caption{Prototype POLAR-2 module where the each scintillator is wrapped with $\SI{12}{\mu m}$ Claryl foils.}
  \label{fig:claryl_proto}
  \hfill
\end{figure}

\newpage
\subsubsection{Thermal Tests}

\begin{figure}[t]
  \centering
  \begin{subfigure}[b]{0.49\textwidth}
     \centering
     \includegraphics[width=1\textwidth]{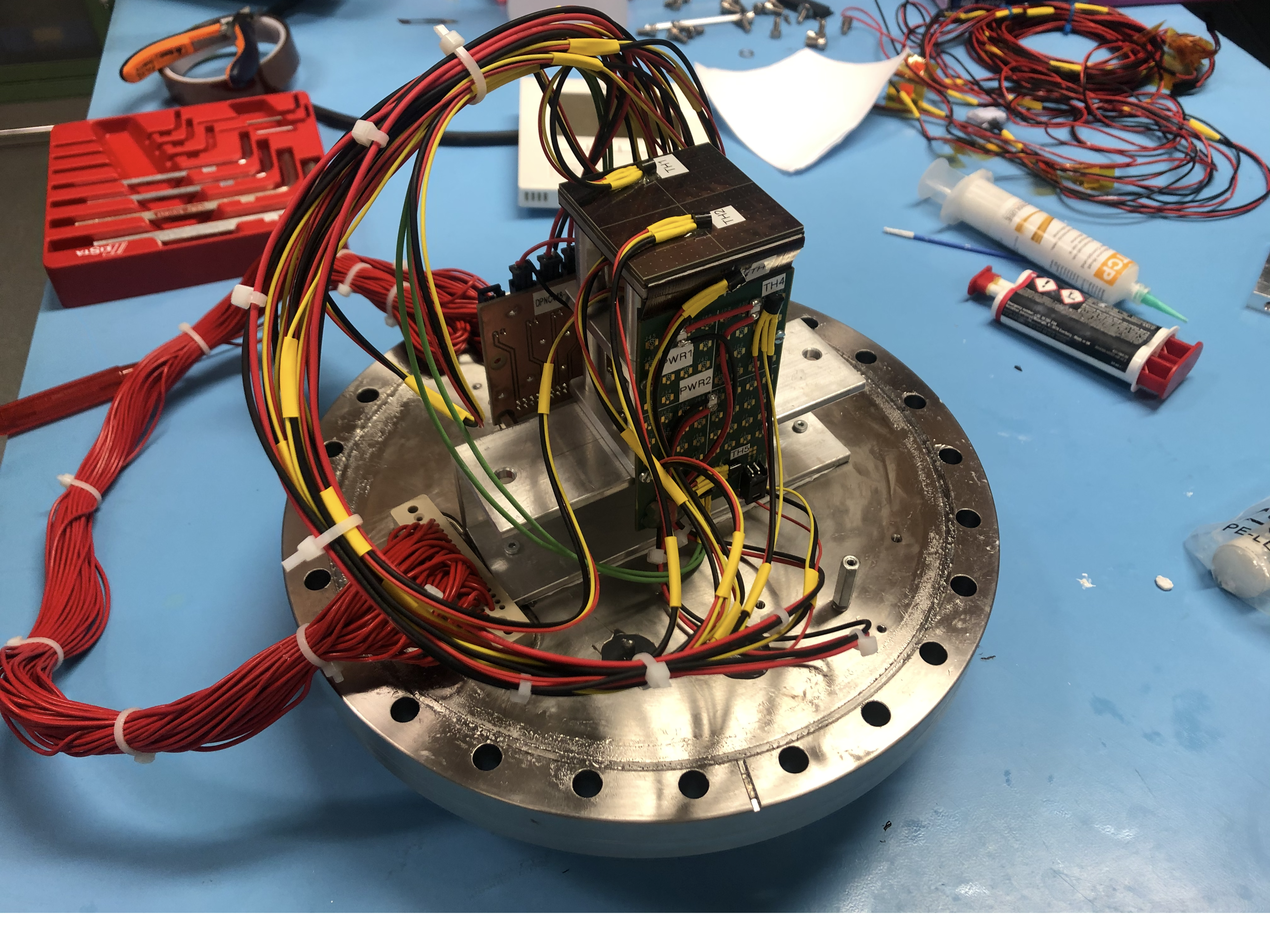}
     \caption{}
     \label{fig:thermal_setup}
  \end{subfigure}
  \hfill
  \begin{subfigure}[b]{0.49\textwidth}
     \centering
     \includegraphics[width=0.5\textwidth]{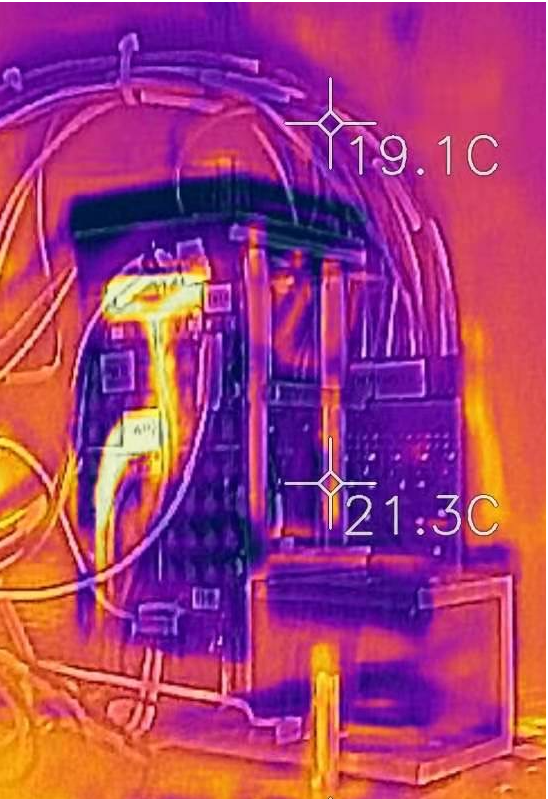}
     \caption{}
     \label{fig:thermal_photo}
  \end{subfigure}
  \caption{}
  \label{fig:thermal}
  \caption{\textbf{a)} Concept mechanical support structure for a POLAR-2 module with a SiPM. Digital heat sensors of type DS18B20 are placed strategicaly around the structure. Three 50$\Omega$ resistors are placed to produce the expected heat produced. \textbf{b)} Thermal photograph of a setup previously cooled to $\SI{0}{^\circ C}$ and then placed at room temperature. Combined, the resistors are consuming $\SI{1.7}{W}$ of power.}
  \hfill
\end{figure}

The electronics of a POLAR-2 module are expected to produce a total of $\SI{1.7}{W}$ ($\SI{1}{W}$ for the FPGA and $\SI{0.7}{W}$ for the CITIROC ASICs). With a power budget of $\approx \SI{2}{W}$ per module, no more than $\approx \SI{0.3}{W}$ should be allocated to the Peltier. At an environmental temperature of $\SI{0}{^\circ C}$ and without any active cooling (only passive), the heat will pass through the SiPMs raising its temperature to about $\SI{4}{^\circ C}$. The preliminary design shown in figure \ref{fig:thermal_setup} and simplistic setup with SiPMs and an FEE. Resistors are placed on the FEE to mimic the heat produced by the components (highlighted by the thermal camera in figure \ref{fig:thermal_photo}). In principle, the structure is held in place by 4 aluminium pillars with a copper tube in its center. This should withstand the vibrations experienced during launch. Copper has a higher heat conductance compared to aluminium and should carry most of the heat away. However, as figure \ref{fig:thermal_photo} already suggests, some of the heat produced also leaks through the aluminium bars. This is not problematic for passive cooling. However, it generates a thermal short circuit when switching on the Peltier element, diminishing its cooling efficiency. \newline
Currently, we are addressing this by replacing aluminium bars with plastic bars. Futhermore, a thermal graphite sheet (Pyrolytic Graphite Sheet - PGS) will be taped from the FEE directly to the copper tube, reducing the heat passing near the SiPM. Thermal simulations are ongoing. Nevertheless, preliminary Peltier tests have shown that with proper heat extraction a temperature difference of $\SI{14}{K}$ could be obtained while only consuming $\SI{0.3}{W}$. \newline

\subsubsection{Overall Performance of Prototype Module}
\label{module_prototype} 

\begin{figure}[t]
  \centering
  \begin{subfigure}[b]{0.49\textwidth}
     \centering
     \includegraphics[width=0.8\textwidth]{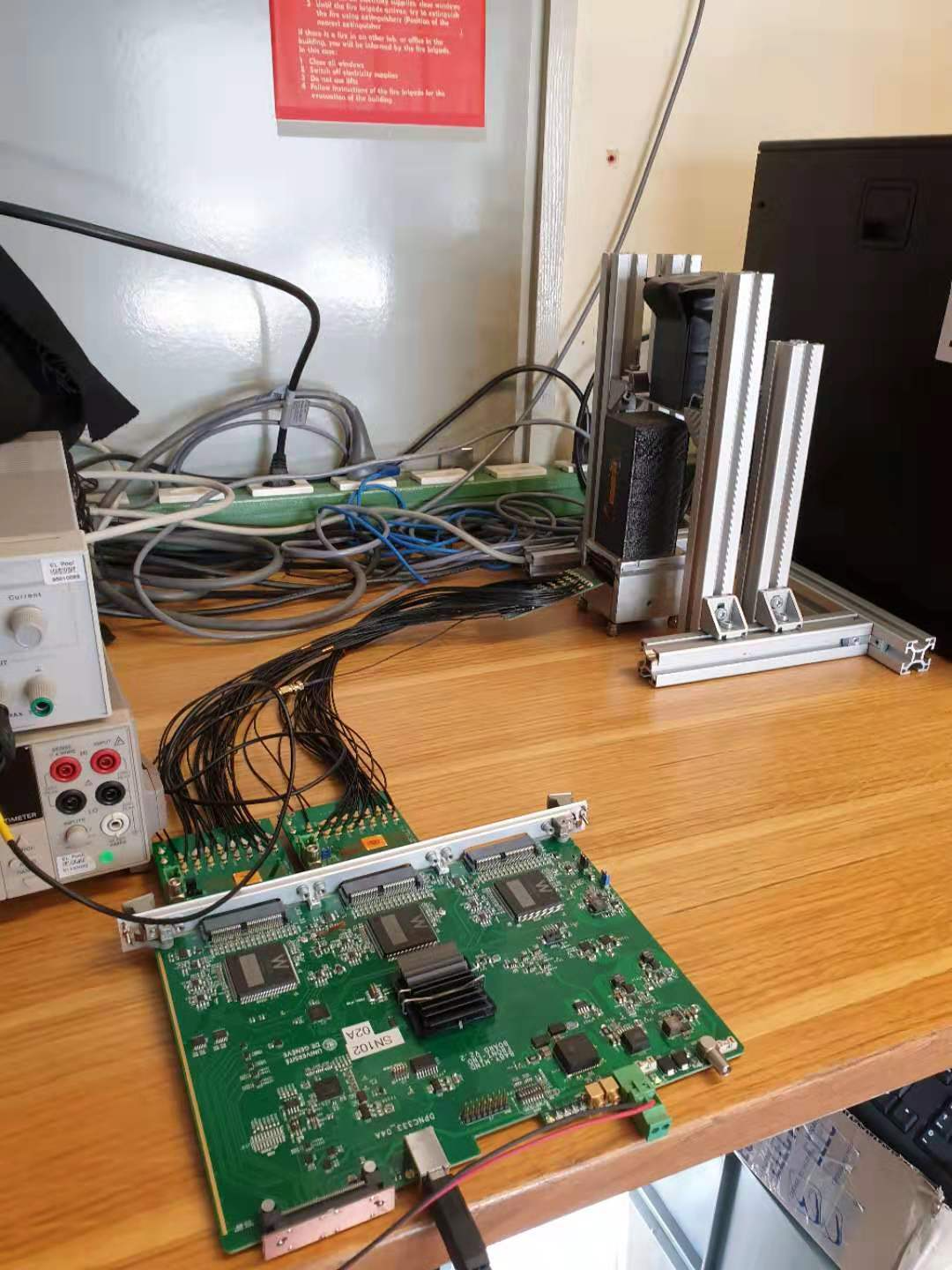}
     \caption{}
     \label{fig:babymind_board}
  \end{subfigure}
  \hfill
  \begin{subfigure}[b]{0.49\textwidth}
     \centering
     \includegraphics[width=0.8\textwidth]{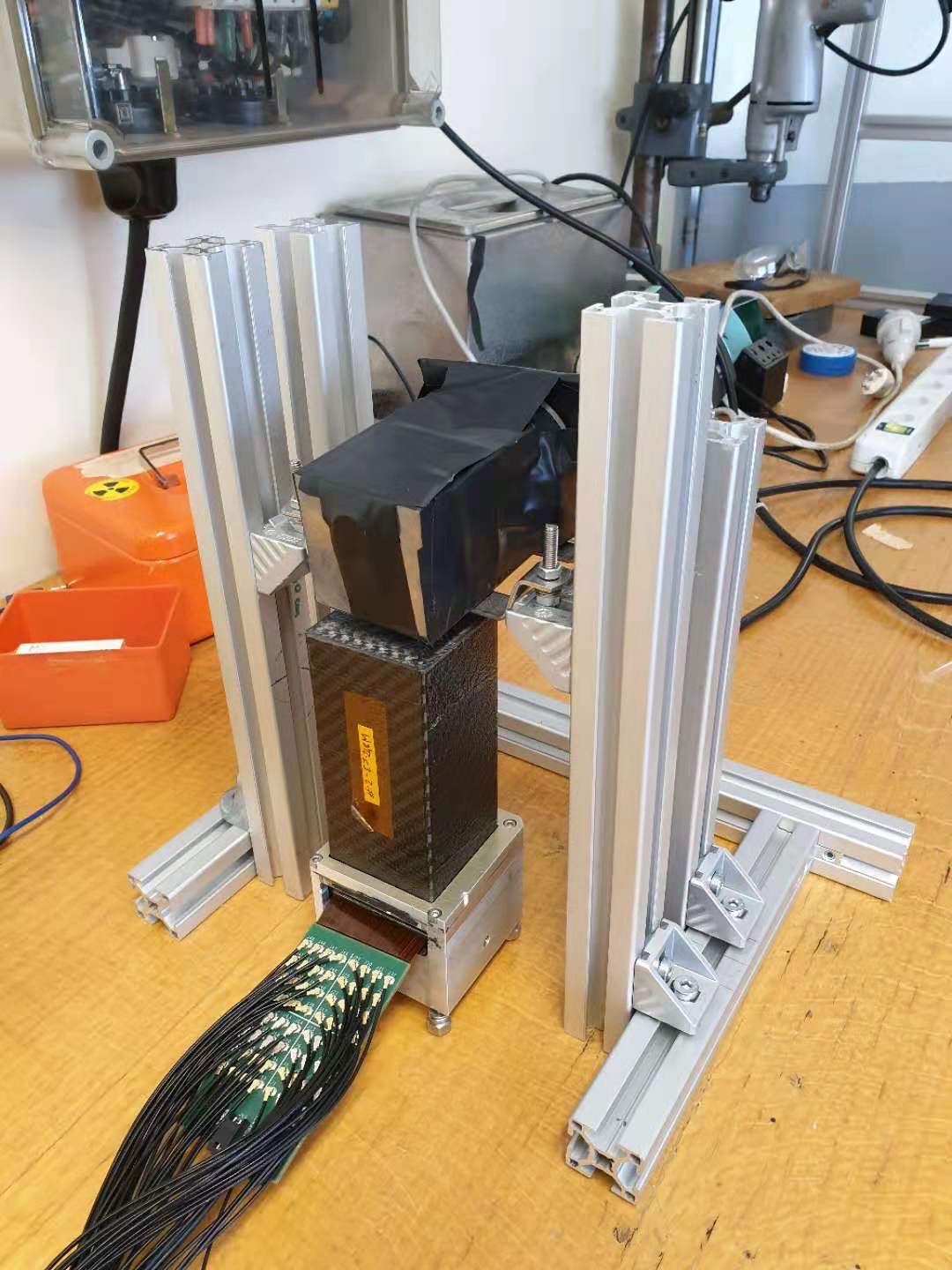}
     \caption{}
     \label{fig:module_cern}
  \end{subfigure}
  \caption{\textbf{a)} SiPM channels read out by two 32-channel CITIROC 1A ASICs on a development board. \textbf{b)} A prototype POLAR-2 , at the CERN facility, irradiated with an Am-241 source.}
  \label{fig:prototype_module}
  \hfill
\end{figure}

At the time of writing, prototype modules are being tested for comparative analyses to the POLAR module (i.e. overall performance) as well as to improve our understanding of the SiPM readout. The prototypes shown in figure \ref{fig:prototype_module} have scintillators composed of EJ-248M. However, they are not truncated as for the POLAR instrument. They are held in place by a plastic grid. Furthermore, each scintillator is individually wrapped by $\SI{12}{\mu m}$ Claryl foils as discussed in section \ref{refl_foil}; their metallic side facing towards the scintillator volume (as seen in figure \ref{fig:module_exposed}). Reflective foil is also placed at one end of the scintillator to prevent photon loss. At the other end, a Hamamatsu S13361-6075NE-04 SiPM is placed, whose data is processed by two 32-channel CITIROC 1A front-end ASICs (illustrated in figure \ref{fig:babymind_board}).\newline

At CERN, the prototype module has been irradiated with an Am-241 source. The radioactive source is placed inside a lead structure which contains a small plastic component ($\approx 1.5\times1.5\times\SI{2.5}{cm}$) to scatter the $\gamma$ particles at an angle of $\SI{90}{^\circ}$ and creating a near 100\% polarized beam. These scattered $\gamma$'s then pass through a collimator before exiting. The lead structure is placed above the module such that the opening of the collimator is directed to the top of the module case (irradiating the scintillators). An illustration of the setup can be seen in figure \ref{fig:module_cern}.\newline
Preliminary results are very encouraging as we see that the cross talk is $<1\%$, significantly less compared to the $\approx15\%$ seen by POLAR modules \cite{Li_2018}. More importantly, the prototype module yields 1 photoelectron/keV, which is already 3x better compared to its predecessor. We remain confident to achieve our targeted 1.5 photoelectrons/keV as there is much room for improvements. These include i) an improved wrapping of the scintillator bars, ii) using scintillators made of EJ-200 and iii) operating at colder temperatures (as these results are obtained at room temperature). Measurements are ongoing; first polarization modulation curves are expected to be produced by the end of 2020. \newline 

\section{Anticipated Scientific Performance}
\label{sec:scientific_performance}

Three different scenarios were compared to gauge the scientific performance of POLAR-2; those being \emph{POLAR}, \emph{POLAR$\times4$} and \emph{POLAR-2}. Results are obtained with the dedicated POLAR MC package discussed in \cite{Kole:2017nhz}. The only difference between POLAR and POLAR$\times4$ is that the number of modules for POLAR$\times4$ has increased from 25 to 100. For POLAR-2, the replacement of MAPMTs by SiPMs was accounted for by increasing the photo-detection efficiency from 0.2 to 0.5 and decreasing the MAPMT/SiPM resolution from 0.6 to 0.1. Furthermore, the optical cross talk has been reduced by an order of magnitude and the threshold is set to 6 photo-electrons \cite{Kole:2020dbs} (, whereas our goal is 4 photo-electrons). The increased light yield by using EJ-200 has not be accounted for. Also, the additional 3.5\% of sensitive volume (discussed in section \ref{refl_foil}) has not been included. \newline

\begin{figure}[h]
  \centering
  \begin{subfigure}[b]{\textwidth}
     \centering
     \includegraphics[width=0.5\textwidth]{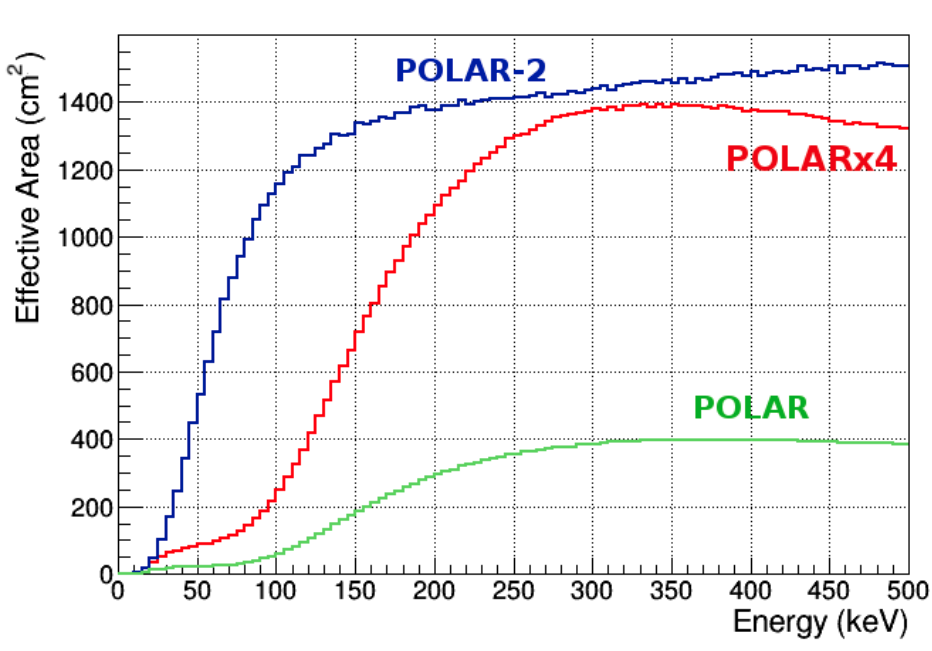}
     \caption{}
     \label{fig:Aeff_polar2}
  \end{subfigure}
  \hfill
  \begin{subfigure}[b]{\textwidth}
     \centering
     \includegraphics[width=0.75\textwidth]{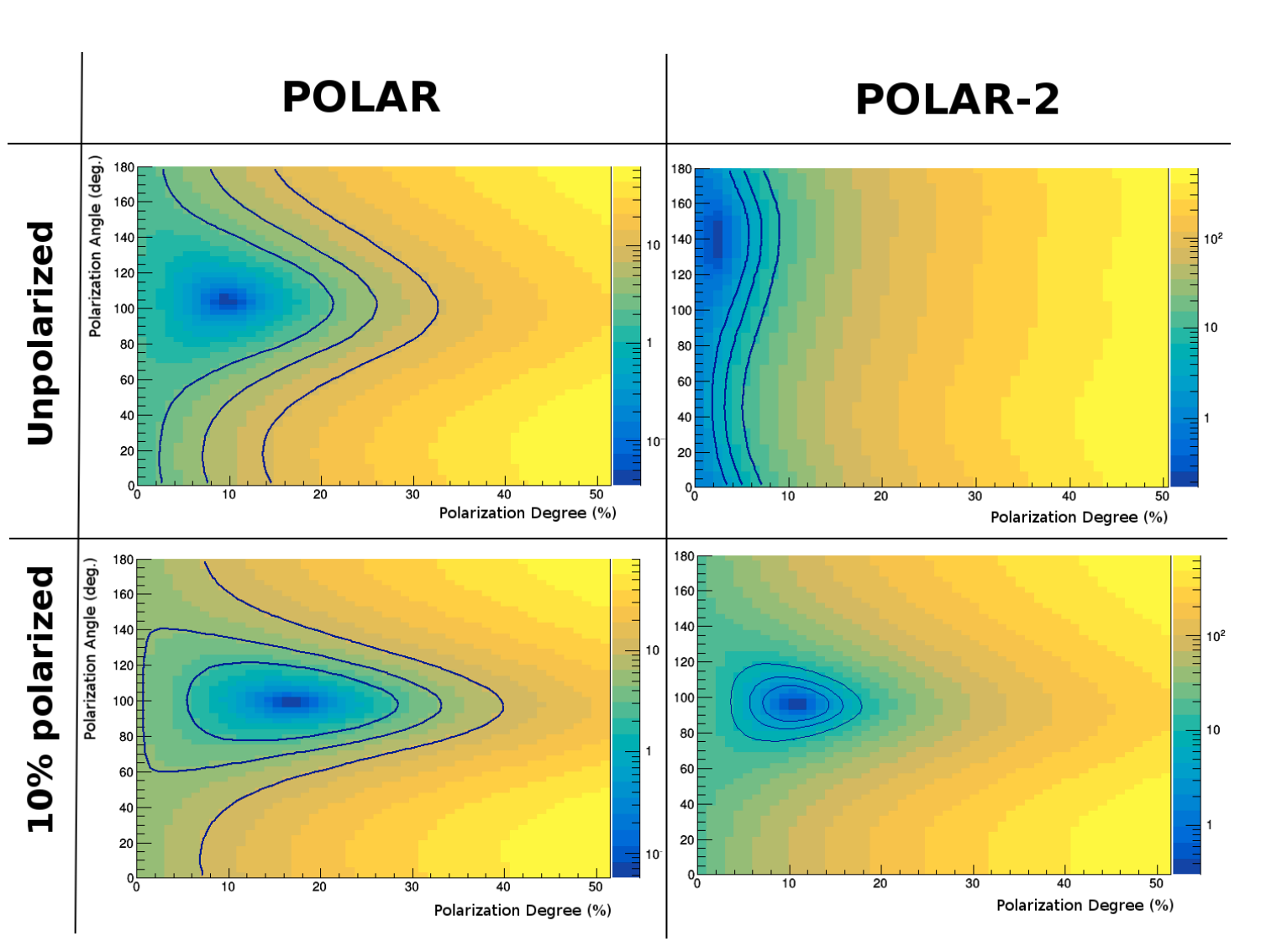}
     \caption{}
     \label{fig:likelihood_polar2}
  \end{subfigure}
  \caption{\textbf{a)} Effective area as function of incoming photon area for POLAR, POLAR$\times4$ (4 times larger than POLAR) and POLAR-2. \textbf{b)} Simulated likelihood distributions for a unpolarized (top) and 10\% polarized (bottom) GRB with a fluence of $\SI{1.3e-5}{\frac{erg}{cm^2}}$. With a POLAR-2 instrument, such polarization measurements can be distinguished \cite{Kole:2020dbs}}
  \label{fig:science}
  \hfill
\end{figure}

A GRB with an incoming angle of $\SI{26}{^\circ}$ off-axis (equal to the \emph{GRB170114A} seen by POLAR \cite{kole2020polar}) is taken as reference to study the effective area and M$_{100}$, which serves as an indicator for the sensitivity of the polarization. Results show (see figure \ref{fig:Aeff_polar2}), that the effective area, especially below $\SI{100}{keV}$, is significantly higher for POLAR-2. This is primarily attributed to the usage of SiPMs. The effective area is also higher for larger energies as a result of the dynamic range of the FEE. For POLAR, the dynamic range is limited to $\approx\SI{300}{keV}$, whereas this can be $\approx\SI{1}{MeV}$ for POLAR-2. No significant changes were found for M$_{100}$. However, it should be noted that the analysis techniques were optimize for the POLAR instrument and therefore not very representative for POLAR-2\cite{Kole:2020dbs}. \emph{GRB170206A}, among the most precisely measured GRBs by POLAR, is used as another reference to investigate the likelihood performance of a simulated event between POLAR and POLAR-2. This is illustrated in figure \ref{fig:likelihood_polar2} where the likelihood distribution of a unpolarized (top) and 10\% polarized (bottom) GRB event. For POLAR, such a GRB event could not be distinguished, whereas POLAR-2 can distinguish this with more than 99\% CL. Further analysis has revealed that POLAR-2 is capable of performing qualitatively similar analyses for GRBs with a fluence of $\SI{2e-6}{\frac{erg}{cm^2}}$ \cite{Kole:2020dbs}. It is therefore projected to measure about 50 GRBs per year with equal or better quality compared to those presented in \cite{Zhang:2019ybt}. \newline

It should be noted that these conclusions are derived from a rather conservative approach towards the POLAR-2 design. We therefore anticipate more improved performances compared to those currently shown in figure \ref{fig:science}. Design changes and calibration data are continuously implemented and updated in a dedicated POLAR-2 MC package (based on \cite{Kole:2017nhz}).

%%%%%%
\newpage

% References
\bibliography{report} % bibliography data in report.bib
\bibliographystyle{spiebib} % makes bibtex use spiebib.bst

\end{document}